\documentclass[aps,twocolumn,PRB,reprint,superscriptaddress]{revtex4}
\usepackage{amsmath}
\usepackage{graphicx}
\usepackage{bm}
\usepackage{amssymb}
\usepackage{hyperref}
\usepackage{dsfont}
\begin{document}
\title{Probing valley filtering effect by Andreev reflection in zigzag graphene nanoribbon}
\author{Kun Luo}
\affiliation{College of Science, Nanjing University of Aeronautics and Astronautics, Nanjing 210016, China}
\author{Tao Zhou}
\affiliation{College of Science, Nanjing University of Aeronautics and Astronautics, Nanjing 210016, China}
\author{Wei Chen}
\email{Corresponding author: weichenphy@nuaa.edu.cn}
\affiliation{College of Science, Nanjing University of Aeronautics and Astronautics, Nanjing 210016, China}

\begin{abstract}
Ballistic point contact (BPC) with zigzag edges in graphene is a main candidate of a valley filter, in which the polarization of the valley degree of freedom can be selected by using a local gate voltage. Here, we propose to detect the valley filtering effect by Andreev reflection. Because electrons in the lowest conduction band and the highest valence band of the BPC possess opposite chirality, the inter-band Andreev reflection is strongly suppressed, after multiple scattering and interference. We draw this conclusion by both the scattering matrix analysis and the numerical simulation. The Andreev reflection as a function of the incident energy of electrons and the local gate voltage at the BPC is obtained, by which the parameter region for a perfect valley filter and the direction of valley polarization can be determined. The Andreev reflection exhibits an oscillatory decay with the length of the BPC, indicating a negative correlation to valley polarization.
\end{abstract}

\maketitle

\section{introduction}
Over the past decade, valleytronics (valley-based electronics) has become a new research field of condensed matter physics \cite{Rycerz,Xiao,Schaibley}. The goal of valleytronics is to manipulate the valley degree of freedom and search for its potential applications in semiconductor technologies and quantum information processing, just analogous to the parallel concepts in spintronics. Great progresses of valleytronics have recently been made because of the emergence of two dimensional materials with honeycomb structures, such as graphene \cite{Novoselov} and monolayer group-VI transition metal dichalcogenides such as MoS$_2$ \cite{Mak,Xu,Liu}. These progresses include the realization of valley polarization \cite{Mak2,Cao,Zeng,Jones}, the discoveries of the valley Hall effect \cite{Xiao,Xiao2,Gorbachev,Sui,Shimazaki,Mak3,Lee}, valley Zeeman effect \cite{Srivastava,Rostami,Aivazian,Stier,MacNeill,Mitioglu} and AC Stark effects \cite{Sie,Kim}, which open a new avenue for the valley-based electronics.

\begin{figure}
\centering
\includegraphics[width=0.48\textwidth]{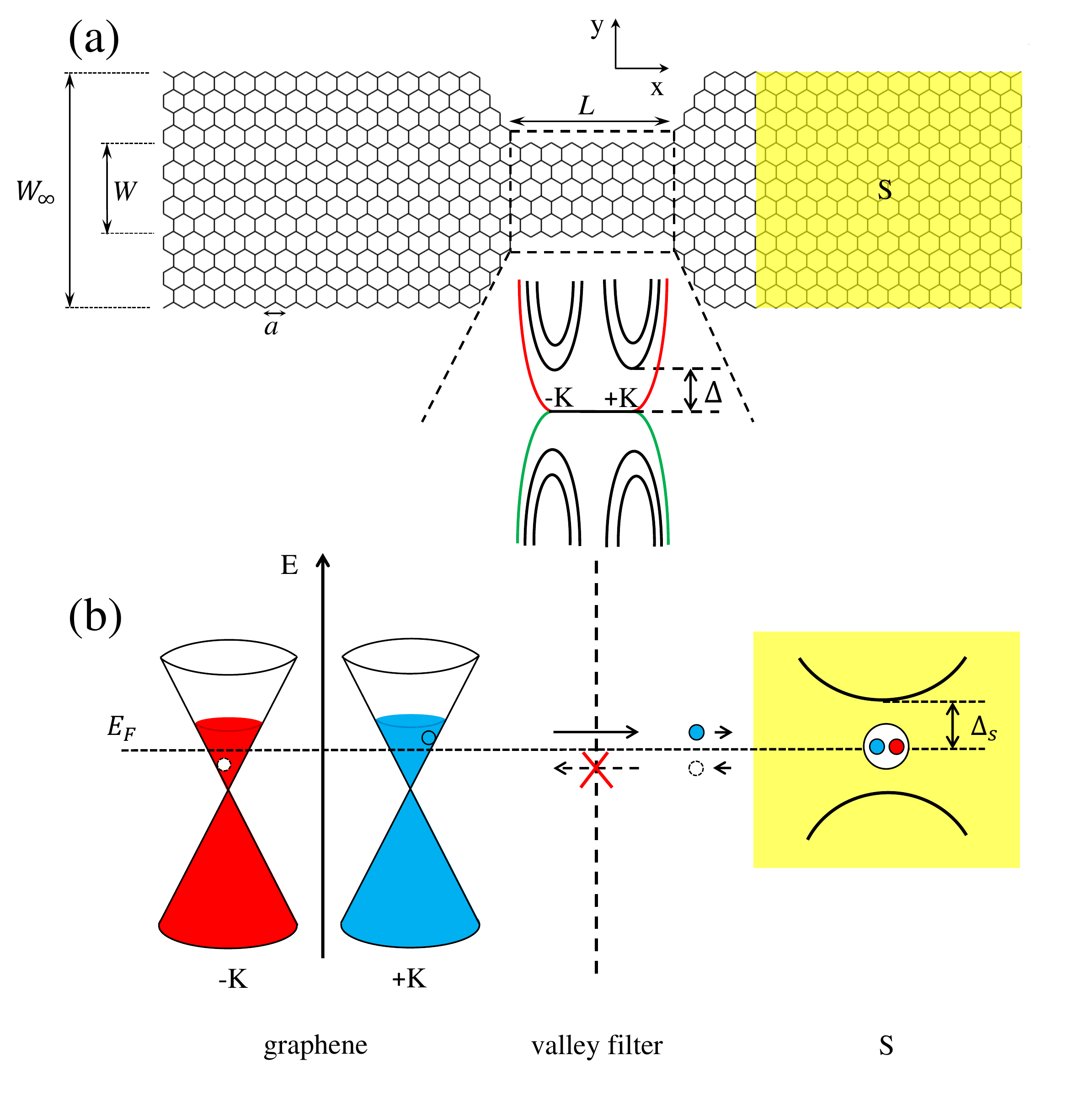}
\caption{(Color online). (a) Schematic of a hybrid junction between a BPC based valley filter and a proximity induced superconducting region denoted by S. The band structure of the BPC is inserted. (b) Illustration of the suppression of the AR. An electron (solid circle) in the conduction band in the $+K$ valley first transmits the valley filter, and then gets Andreev reflected as a hole (open circle) in the valence band in the $-K$ valley. The hole in the $-K$ valley cannot transmit the BPC, and thus the AR is suppressed.
 } \label{fig1}
\end{figure}

Just like the spin filter in spintronics, valley filter is a central component of valleytronics, which plays a central role in generation and detection of valley polarized current \cite{Rycerz}. Various proposals for valley filter in graphene have been put forward, by fabricating ballistic point contact (BPC) with zigzag edges \cite{Rycerz}, line defect \cite{Gunlycke,Cheng}, topological domain wall \cite{Pan}, electrostatic potentials in bilayer graphene \cite{Costa}, nanobubbles \cite{Settnes}, polycrystalline graphene \cite{Nguyen}, and using strain \cite{Peeters}. However, to the best of our knowledge, experimental detection of the valley polarized current generated by the valley filter remains challenging. Several proposals of using optical method to generate and probe valley polarization are put forward \cite{Golub,Golub2,Wehling}, but the imbalance in valley populations is different from a valley polarized current generated by the valley filter. The valley Hall effect may achieve the purpose \cite{Gorbachev,Shimazaki}, while an integration of valley filter and valley Hall measurement still needs more efforts. In this work, we propose to detect the valley polarization of current by measuring Andreev reflection (AR) in the valley filter-superconductor junction, as shown in Fig. \ref{fig1}(a). The valley filter here is specified to the BPC in the zigzag graphene nanoribbon. Within the BPC, the lowest conduction band and the highest valence band dominate electron transport, which is sketched by the inset in Fig. \ref{fig1}(a). Outside the flat band, dispersive electrons in the conduction and valence bands possess opposite chirality, and can be described by the Hamiltonian $h=\pm \hbar v_0\delta k\tau_z$, where the small wave vector $\delta k$ measured from the Dirac points takes positive and negative values for the $+K$ and $-K$ valleys, respectively. The Pauli matrix $\tau_z$ operates in the valley degree of freedom, and $v_0$ is the velocity. Because the conduction and valence bands possess opposite style of valley-momentum locking, the polarization of the valley current can be conveniently inverted by imposing a local gate voltage to the BPC \cite{Rycerz}.

Here, we show that the remarkable function of the BPC valley filter, that is generating high-purity valley current and selecting its polarization by the gate voltage, can be exactly reflected by the Andreev spectra \cite{Akhmerov}. When the graphene sheet contacts with an s-wave superconductor, the proximity effect induces electron pairing with zero net momentum. Thus the pairing occurs between $+K$ and $-K$ valleys in graphene \cite{Beenakker,Beenakker2}, while the opposite chirality forbids the AR between the conduction and valence bands \cite{Rainis,Bovenzi,Wang}. Using the scattering matrix analysis, we prove that the suppression of AR for highly polarized valley current survives after multiple scattering between the BPC and the graphene-superconductor (GS) interfaces. The conclusion is supported by rigorous numerical simulation based on the tight-binding model. The AR probability as a function of the incident energy of electrons and the gate voltage at the BPC is calculated, by which the parameter region for a perfect valley filter can be determined. We also investigate the dependence of AR on the length of the BPC, which shows an oscillatory decay behavior, indicating its negative correlation to the valley polarization. Our work is different from the previous one on the AR in a translation invariant graphene nanoribbon in Ref. \cite{Rainis}, where the effect of the valley filter is not included, so that the relation between the valley polarization and AR probability cannot be extracted.

The rest of this paper is organized as follows. In Sec. II, we present a scattering matrix description of the AR and prove that the BPC based valley filter results in suppression of AR. In order to give rigorous results, numerical simulation is carried out to investigate the relation between valley polarization and AR probability in Sec. III. Finally, a brief summery and prospect are given in Sec. IV.

\section{scattering matrix analysis of Andreev reflection}\label{sec1}

To reveal the underlying mechanism of the suppression of AR by the BPC valley filter, we first adopt an analytic description of the problem by the method of scattering matrix. The whole setup is sketched in Fig. \ref{fig1}(a), where a valley filter is constructed by a BPC structure in a zigzag graphene nanoribbon. At the right side of the nanoribbon, an s-wave superconductor is deposited, which induces pair potential $\Delta_s$ in graphene due to the proximity effect \cite{Gennes,Beenakker,Beenakker2}. The valley filtering effect occurs in the lowest conduction band and the highest valence band of the BPC, which are concerned here. The energy spacing between them and the neighboring transverse modes is $\Delta=3\pi\hbar v_0/2W$ \cite{Rycerz} as shown in Fig. \ref{fig1}(a), where $W$ is the width of the BPC. Because the BPC is narrow, the mode spacing is much larger than the superconducting gap, that is $\Delta\gg\Delta_s$. The suppression of AR occurs when the incident electron and the reflected hole transport through the conduction and valence bands of the BPC, respectively. In order to draw our main conclusion, we first assume that the Fermi energy coincides with the flat band in the BPC in Fig. \ref{fig1}(a), so that the electron and hole definitely transmit the BPC through different bands. In the wide regions outside the BPC, electrostatic potential is applied such that the Fermi level passes through $2N+1$ transverse modes in the conduction bands.

The scattering at two interfaces, i.e., the BPC region and the GS interface can be described by the scattering matrices. The coefficients of the incident and reflected waves at the BPC are related by $(b_L,b_R)^T=s_1(a_L,a_R)^T$, with $s_1$ being expressed by
\begin{equation}\label{s10}
s_1=\left(
          \begin{array}{cc}
            r & t' \\
            t & r' \\
          \end{array}
        \right)
\end{equation}
Here the incident wave from the left side is expressed as $a_L=(a_1,...,a_{2N+1};a_1^h,...,a_{2N+1}^h)^T$, where there are $2N+1$ transverse modes for both electron and hole components. The electron components $a_i$ are arranged such that a lower index corresponds to the incident states with a larger momentum. This means the transverse modes $1, 2,...,N+1$ locate in the $+K$ valley, while the remaining modes belong to the $-K$ valley. The outgoing modes are the time reversal of the incident ones as $T\psi_{\text{in}}(k_i,E)=\psi_{\text{out}}(-k_i,E)$, which means that for an incident wave $a_{i}$ with the wave vector $k_i$, the reflection wave $b_{i}$ has a wave vector $-k_i$. The incident hole waves $a_i^h$ are related to the electron ones $a_i$ through the particle-hole transformation as $\Xi\psi_{\text{in}}(k_i,E)=\psi_{\text{in}}^h(-k_i,-E)$, where the superior $h$ denotes the hole component and the energy $E$ is measured from the Fermi energy. The similar relation can be found between the outgoing electron states $b_i$ and hole states $b_i^h$.

For the BPC valley filter, time reversal symmetry requires that the scattering matrix is symmetric \cite{Datta}, that is $t'=t^T, r^T=r, r'^T=r'$. Moreover, the valley filter also possesses inversion symmetry about the BPC center along the $x$ direction, which further imposes the constraints $r'=r$ and $t'=t$ to the scattering matrix \cite{Merzbacher}. As a result, the scattering matrix in Eq. (\ref{s10}) reduces to
\begin{equation}
s_1=\left(
          \begin{array}{cc}
            r & t \\
            t & r \\
          \end{array}
        \right)
\end{equation}
with the submatrices satisfying $t^T=t, r^T=r$. Current conservation law requires the scattering matrix to be unitary, that means $r=-(t^*)^{-1}r^*t$ and $|r|^2+|t|^2=\mathds{1}$, with $\mathds{1}$ being the unit matrix.

In the normal region without superconducting pair potential, the electron and hole elements are decoupled, so that the scattering matrix is block diagonal. The reflection submatrix can be defined as
\begin{equation}
r(E)=\left(
    \begin{array}{cc}
      r_e(E) & 0 \\
      0 & r_h(E) \\
    \end{array}
  \right).
\end{equation}
The electron part is expressed as
\begin{equation}
r_e(E)=\left(
      \begin{array}{cc}
        r_{-+} & r_{--} \\
        r_{++} & r_{+-} \\
      \end{array}
    \right)
\end{equation}
where $r_{\alpha\beta}$ describes the scattering from valley $\beta$ to valley $\alpha$, where the subscripts $\alpha,\beta=\pm$ denote $\pm K$ valleys respectively.
The hole part of the reflection matrix can be obtained by using the particle-hole symmetry of the Bogoliubov-de Gennes (BdG) Hamiltonian \cite{Gennes}, which is
\begin{equation}
r_h(E)=\left(
      \begin{array}{cc}
        r^h_{+-} & r^h_{++} \\
        r^h_{--} & r^h_{-+} \\
      \end{array}
    \right)=r_e^*(-E).
\end{equation}
Note that creation of a hole excitation is equivalent to annihilation of an electron with opposite momentum and energy, so that the valley subscripts of the hole matrix are opposite to that for the electron.

Similarly, for the transmission submatrix, we write
\begin{equation}
t(E)=\left(
    \begin{array}{cc}
      t_e(E) & 0 \\
      0 & t_h(E) \\
    \end{array}
  \right),
\end{equation}
where the electron part can be expressed as
\begin{equation}\label{te}
t_e(E)=\left(
      \begin{array}{cc}
        t_{++} & t_{+-} \\
        t_{-+} & t_{--} \\
      \end{array}
    \right),
\end{equation}
and the corresponding hole part is
\begin{equation}\label{th}
t_h(E)=\left(
      \begin{array}{cc}
        t^h_{--} & t^h_{-+} \\
        t^h_{+-} & t^h_{++} \\
      \end{array}
    \right)=t_e^*(-E).
\end{equation}

As the electron transmits the BPC, it gets scattered at the GS interface. For an incident energy below the superconducting gap, that is $0\leq E<\Delta$, only reflection processes occur, which can be described by the reflection matrix as $a_R=s_2 b_R$, with $s_2$ taking the form of
\begin{equation}
s_2=\left(
      \begin{array}{cc}
        r_n(E) & r_a(E) \\
        r_a^*(-E) & r_n^*(-E) \\
      \end{array}
    \right),
\end{equation}
where $r_a,r_n$ refer to AR and normal reflection respectively. Note that for the GS interface, $b_R$ and $a_R$ become the incident and outgoing waves respectively. As the chemical potentials in the normal and superconducting regions are equal, the normal reflection is zero, that is $r_n=0$. Moreover, because the energy difference between different transverse modes of the BPC is $\Delta\simeq 280meV$, while the gap of the superconductor is $\Delta_s\simeq1meV$, the AR between different transverse modes can be neglected. Thus the submatrices of AR can be simply described by the unit matrix as $r_a=\mathds{1}$, and $s_2$ reduces to
\begin{equation}
s_2=\left(
      \begin{array}{cc}
        0 & \mathds{1} \\
        \mathds{1} & 0 \\
      \end{array}
    \right).
\end{equation}

Below the superconducting gap, quasi-particle cannot survive in the superconducting region, so that the incident particle coming from the left side is completely  reflected. This can be denoted by $b_L=\mathcal{R} a_L$, where the final reflection matrix $\mathcal{R}$ can be obtained by combining $s_1$ and $s_2$, resulting in \cite{Chen}
\begin{equation}
\mathcal{R}=r+t s_2(\mathds{1}-r s_2)^{-1} t.
\end{equation}
In order to obtain the final amplitude of AR, we expand the inverse matrix term as $(\mathds{1}-r s_2)^{-1}=\sum_{n=0}(r s_2)^n$. It can be verified that the odd and even terms of $rs_2$ contain only off-diagonal and diagonal elements respectively.
Therefore, only even terms contribute to the final AR due to the perfect AR at the GS interface. This corresponds to the off-diagonal elements of $\mathcal{R}$, which describe the electron-hole scattering. The submatrix $\mathcal{R}_{he}$ takes the form of
\begin{equation}\label{Rhe}
\mathcal{R}_{he}=t_h\tilde{r}t_e,
\end{equation}
where $\tilde{r}=(\mathds{1}-r_er_h)^{-1}$.

The suppression of AR by the valley filter can be extracted from the above expression. For a perfect valley filter, most elements in $t_e(E)$ in Eq. (\ref{te}) is negligibly small. In the special case that the Fermi energy coincides with the flat band in the BPC, for an incident electron with positive energy $E>0$, the only nonzero submatrix is $t_{++}$. For the other submatrices, the transmission into $-K$ valley denoted by $t_{-+}$ and $t_{--}$ are completely filtered by BPS and the elements in $t_{+-}$ are also negligible due to the large momentum mismatch between $\pm K$ valleys. Similarly, the transmission for hole is dominated by $t^h_{++}(E)=t^*_{--}(-E)$. Because of the particle-hole symmetry of the matrix in Eq. (\ref{th}), the scattering matrix for hole equals that for electron with opposite energy and valley up to a complex conjugate. Based on the argument above, we conclude that the remaining block in Eq. (\ref{Rhe}) is picked out by the non-vanishing parts in $t_h$ and $t_e$, that is $\mathcal{R}_{he}^{21}=t^*_{--}(-E)\tilde{r}_{21}(E)t_{++}(E)$, with $\tilde{r}_{21}=\sum_{n=1}(r_{++}r^h_{+-}+r_{+-}r^h_{--})^n$. Furthermore, it can be verified numerically that the probabilities for the inter-valley reflection are much smaller than that for the intra-valley one, that is $\text{Tr}(r_{\alpha\bar{\alpha}}^\dag r_{\alpha\bar{\alpha}})\ll \text{Tr}(r_{\alpha\alpha}^\dag r_{\alpha\alpha})$, with $\bar{\alpha}$ denoting an opposite valley of $\alpha$. In this case, the reflection matrix is of the order $\tilde{r}_{21}\sim r_{+-}$, which indicates a small reflection probability. Based on the estimation above, we conclude that the AR probability given by $R_A=\text{Tr}(\mathcal{R}_{he}^\dag \mathcal{R}_{he})$ is strongly suppressed. For more general cases, the Fermi energy can deviate from the flat band in the BPC by a value of $U_0$, which can be tuned by the gate voltage. The suppression of AR still holds as long as the incident electron and the reflected hole transport through different bands of the BPC, which means the incident energy satisfies $|E|>|U_0|$. Otherwise, nearly perfect AR will occur, and its probability $R_A$ approaches unity, corresponding to the single transverse mode available in the BPC.

From the above discussion, it can be seen that the suppression of AR is a direct result of the valley filtering effect of the BPC, in which the conduction and valence bands select electrons with opposite valley degree of freedom. Besides, the weak intra-valley reflection at the BPC is also a precondition for the suppression. Otherwise, it cannot survive after multiple scattering and interference between the BPC and the GS interface. The scenario here is different from the case in half-metallic material, where the AR is suppressed by the imbalance of the populations of electrons with opposite spins at the Fermi energy \cite{Soulen,Ji}.

\section{Numerical simulation}

\begin{figure}
\centering
\includegraphics[width=0.48\textwidth]{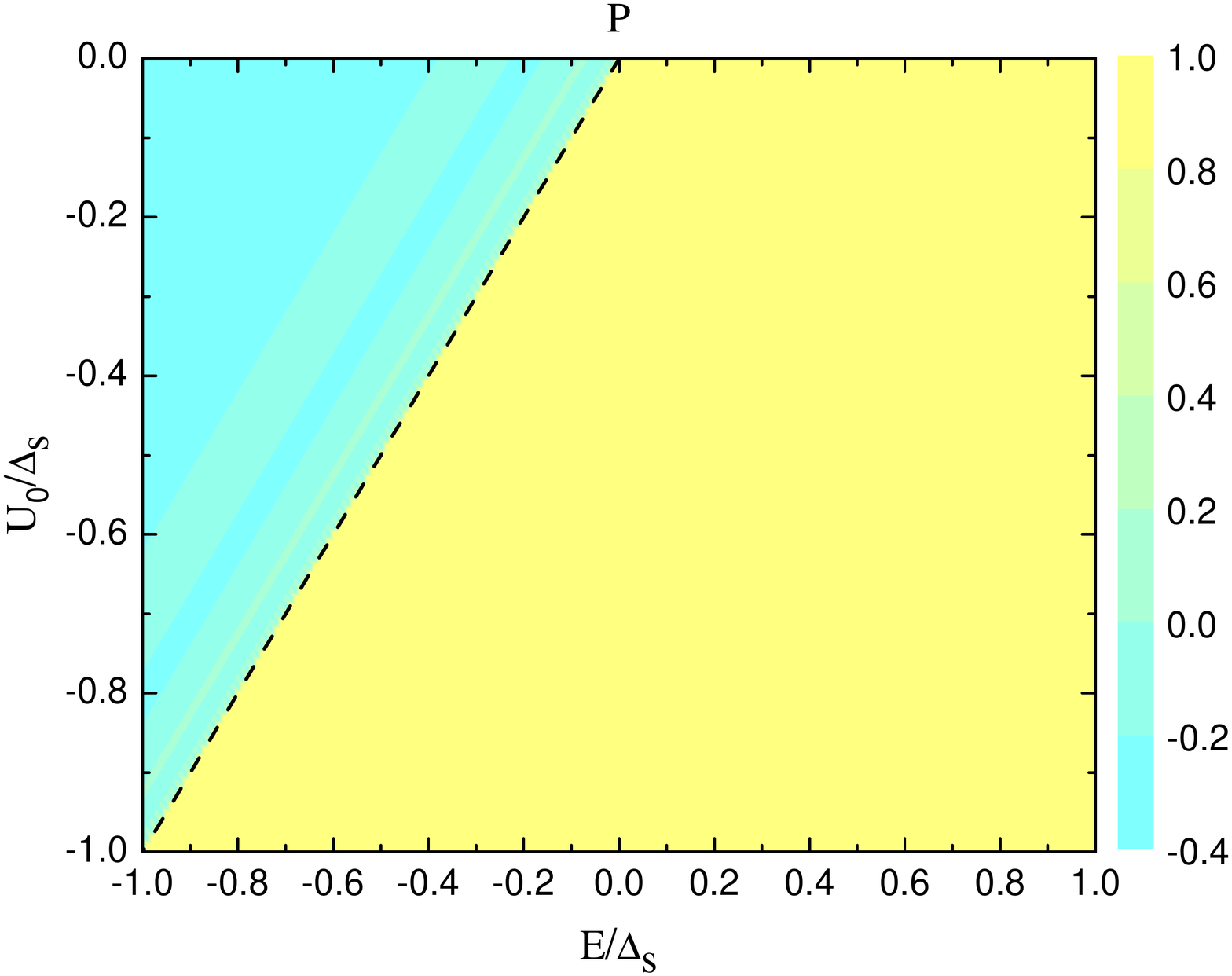}
\caption{(Color online). Plot of the valley polarization $P$ as a function of potential $U_0$ and incident energy $E$. The dashed line denotes the boundary between opposite valley polarization.}\label{fig2}
\end{figure}

Based on the scattering matrix analysis, we have draw the qualitative conclusion that the AR gets suppressed by the BPC valley filter. In the following, we perform numerical simulation based on the lattice model to give rigorous results to support this conclusion. The tight-binding Hamiltonian in graphene is
\begin{equation}\label{H}
H=-t\sum_{\langle i,j\rangle} (a^\dag_{i}a_{j} + H.c.) + \sum_{i} U_{i}a^\dag_{i}a_{i},
\end{equation}
where $a_{i}$ is the Fermi operator at the site $i$, $t\approx 2.8$ eV is the nearest-neighbor hopping energy and $U_i$ is the on-site potential energy. As a superconductor is included, the system is described by the BdG Hamiltonian \cite{Gennes} as
\begin{equation}\label{1}
H_{\text{BdG}}=\left(
\begin{array}{cc}
H-E_F & \Delta_s(x) \\
\Delta_s^*(x) & E_F-\mathcal{T}H\mathcal{T}^{-1} \\
\end{array}
\right),
\end{equation}
where $H$ is the single-particle Hamiltonian (\ref{H}), $E_F$ is the Fermi energy, $\Delta_s(x)$ is the superconducting pair potential, and $\mathcal{T}$ is the time-reversal operator. The pair potential is set as $\Delta_s(x)=1$ meV for the superconducting region and vanishes in the normal region.

As sketched in Fig. \ref{fig1}(a), the geometric parameters of the setup are set as follows:
the length of the BPC is $L=52a$, and its width is $W=20\sqrt{3}a$, with $a$ being the lattice constant; the width of the wide region of the valley filter and the superconducting region is $W_{\infty}=70\sqrt{3}a$. The on-site potential $U_{i}$ is set to $U_0$ for the BPC region and to $U_\infty$ elsewhere. In the numerical calculation, $U_{\infty}=-t/3$ is adopted and $U_0$ varies in the energy scale of $\Delta_s$, which can be tuned by the gate voltage. The numerical calculation is performed by using the Kwant package \cite{kwant}.

First we calculate the valley polarization $P$ for the BPC valley filter without the superconductor, which is defined by
\begin{equation}
P=\sum_{\alpha,\beta=\pm}\alpha\text{Tr}( t_{\alpha\beta}^\dag t_{\alpha\beta})/\text{Tr}(t_e^\dag t_e),
\end{equation}
where the matrices $t_{\alpha\beta}$ are defined in Eq. (\ref{te}). This expression can be further simplified to $P=\text{Tr}(t_{++}^\dag t_{++}-t_{--}^\dag t_{--})/\text{Tr}(t_e^\dag t_e)$, according to the condition $t^T=t$ restricted by the symmetry.
The valley polarization as a function of the incident energy $E$ of the electron and the gate voltage $U_0$ is shown in Fig. \ref{fig2}. There is a clear boundary between opposite valley polarization, that is $E=U_0$, which indicates the incident energy coincides with the flat band of the BPC. As $E>U_0$, the valley polarization $P\simeq1$, because that only the channel at $+K$ valley is available in the BPC. Oppositely, as $E<U_0$, the valley polarization takes negative values, corresponding to a valence band dominated transport in the BPC. One can see that the negative valley polarization cannot reach its saturation value of $-1$ within the energy scale of $\Delta_s$. This stems from the considerable mismatch of the momentums around the $-K$ valley for the incident states and the propagation channel inside the BPC.

\begin{figure}
\centering
\includegraphics[width=0.48\textwidth]{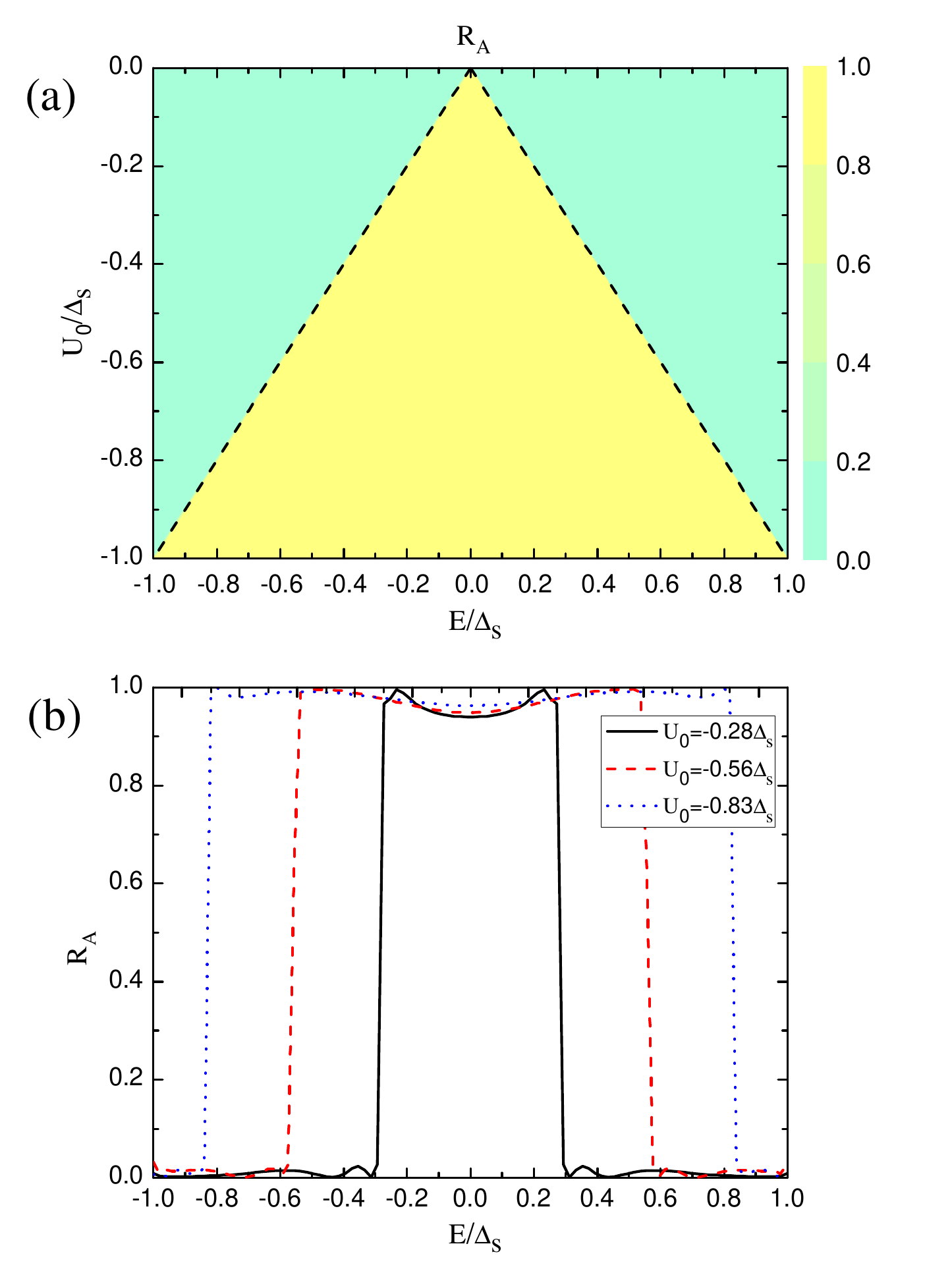}
\caption{(Color online). (a) Plot of AR probability $R_A$ as a function of potential $U_0$ and incident energy $E$. The dashed line divides the whole region into two parts, with perfect AR and suppressed AR. (b) Plot of $R_A$ as a function of $E$ for fixed values of $U_0$.}\label{fig3}
\end{figure}

Next we introduce the superconductor to the system as shown in Fig. \ref{fig1}(a) and calculate the AR probability $R_A$ for electrons incident from the left side. The AR probability as a function of $E$ and $U_0$ is shown in Fig. \ref{fig3}(a). The parameter regions for nearly perfect AR and strongly suppressed AR is divided by the line segment $|E|=|U_0|$, as is expected in the analytical estimation in the previous section. As $|E|>|U_0|$, the incident electron and reflected hole propagate through the conduction and valence bands in the BPC respectively. In this case, the opposite chirality prevents the AR process, even though multiple scattering exists between the BPC and the GS interface. When $|E|<|U_0|$, nearly perfect AR occurs, in which case the incident and reflected hole propagate in the same band inside the BPC. We apparently show the values of AR probability varying with the incident energy in Fig. \ref{fig3}(b) by keeping $U_0$ a constant. One can see a transition between weak and nearly perfect AR occurs as the incident energy crosses the boundary of $|E|=|U_0|$. Given that perfect AR takes place at the GS interface, we conclude that the suppression of AR completely stems from the BPC induced valley polarization. As a result, the Andreev spectra can be used to probe the valley polarization.

\begin{figure}
\centering
\includegraphics[width=0.48\textwidth]{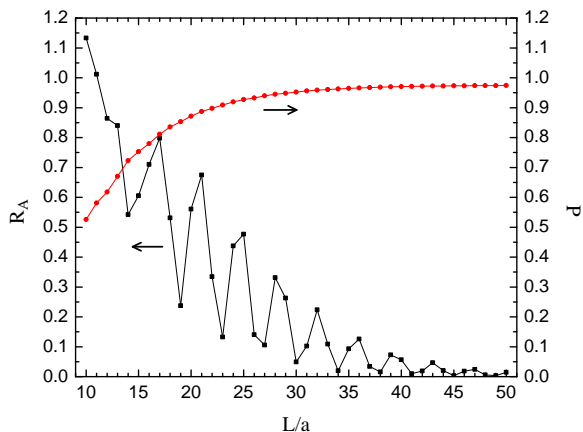}\\
\caption{(Color online). Plot of the AR probability $R_A$ and valley polarization $P$ as a function of the length $L$ of the BPC, with an incident energy $E=0.5\Delta_s$.}\label{fig4}
\end{figure}

It has been verified that for a perfect valley filter, where the valley polarization $P$ of the current approaches unity, the AR is strongly suppressed when $|E|>|U_0|$. Next we investigate the AR for valley filters with different valley polarizations, which can be achieved by changing the length $L$ of the BPC. In Fig. \ref{fig4}, we plot the dependence of valley polarization $P$ and AR probability $R_A$ on the length of the BPC. As $L$ varies from $10a$ to $50a$, $P$ increases monotonically from 0.52 to its saturation value approaching 1. On the contrary, $R_A$ exhibits an oscillatory decay. Apart from the interference induced oscillatory, the AR probability exhibits a negative correlation to the valley polarization. As valley polarization approaches unity, it can definitely be detected by the suppression of AR process.

The AR probability within the superconducting gap $\Delta_s$ can be directly measured by the differential conductance through $G(eV)=2R_A(eV)e^2/h$ at a bias voltage $eV$. Comparing Fig. \ref{fig2} and \ref{fig3}(a), one can see that the boundary of opposite valley polarization coincides with one of the lines dividing strong and weak AR, that is $E=U_0$. As a result, the parameter distribution of valley polarization can be determined by the Andreev spectra. It is beneficial for experiment that a switching behavior of the AR probability occurs as the incident energy crosses the flat band. Such a clear signal means that the suppression of AR is directly caused by the valley filter. Otherwise, if the AR probability changes a little in the whole energy interval, then it is difficult to draw the key mechanism for the AR suppression. For experimental realization of our proposal, we note that the state of the art of the graphene engineering strongly shows the potential to construct the BPC valley filter in the near future \cite{Jiao,Kosynkin,Cai,Novoselov2,Mohanty,Chen}, and recent progress on specular AR in graphene has also confirmed the superconducting proximity effect in graphene \cite{Efetov}. Therefore, it is reasonable to anticipate the realization of our proposal.

\section{summary and outlook}
In summary, we investigate the AR process in the BPC valley filter constructed on the zigzag graphene nanoribbon. We focus on the relation between the AR and the valley polarization both analytically and numerically. It is found that the valley polarization leads to a suppression of AR. According to the Andreev spectra, the parameter region of high valley polarization and the direction of the valley polarization can be determined. Our work provide a promising and effective way to confirm the function of the valley filter. Although our study is based on the BPC valley filter, the idea of detecting valley polarized current by AR can be hopefully extended to other schemes of valley filter \cite{Gunlycke,Cheng,Pan,Costa,Settnes,Nguyen,Peeters}. This is reasonable by noting that the electron pairing occurs between $\pm K$ valleys, while the valley filter filters one of them.

\begin{acknowledgments}
We would like to thank B. G. Wang for stimulating this research and sharing his insight on the close relation between valley filter and Andreev reflection. This work was supported by the National Natural Science Foundation of China under Grant No. 11504171 and No. 11374005 and the Natural Science Foundation of Jiangsu Province in China under Grants No. BK20150734.
\end{acknowledgments}

\end{document}